\documentclass{elsart}

\usepackage{natbib}
\usepackage{graphicx}

\begin{document}

\begin{frontmatter}



\title{Low-temperature dynamics of spin glasses: Walking in the energy landscape}


\author{J. Krawczyk}, 
\ead{jarek@theory.phy.tu-dresden.de} 
\author{S. Kobe} 
\ead{Kobe@physik.tu-dresden.de}
\address{Institut f\"ur Theoretische Physik, Technische Universit\"at Dresden, \\ D-01062 Dresden, Germany}
\begin{abstract}
We analyse the relationship between dynamics and conf\/iguration space structure of Ising spin glass systems. 
The exact knowledge of the structure of the low--energy landscape is used to study the relaxation  of the 
system by random walk in the conf\/iguration space. The inf\/luence of the size of the valleys, clusters and energy barriers 
and the connectivity between them on the spin correlation function is shown. 
\end{abstract}


\end{frontmatter}


\section{Introduction}

In general, systems which comprehend disorder and competing interactions (frustration) are characterised by a complex landscape
in the high dimensional conf\/iguration space. The dynamics of such systems is strongly correlated with their complex topography 
of the phase space. 
Consequently, dynamical processes are determined by the movement in the space. The strong increase of the relaxation
time for low temperatures is related to metastable states and global minima acting as basins of attractions. It has been established that 
the underlying mechanism is uniform for dif\/ferent systems, e.g. spin glasses, supercooled liquids  and the protein folding problem.
The physical understanding of the behaviour of these systems from a microscopic point of view is a major challenge. It would demand 
the knowledge of the huge number of system states, the connectivity of these states in the conf\/iguration space and their correlation
with real space properties. Numerical investigations are restricted to small systems and various procedures are proposed, e.g., molecular
dynamics simulation presented for supercooled liquids \cite{bu_heu,Scir}, pocket analysis of the phase space around a local minima \cite{sib_sch},
and random walk in the energy landscape of spin glasses \cite{kl_kobe,glt}. In this work we calculate the exact low--energy landscape
for a f\/inite spin glass system and  study the dynamics  by a random walk through the  conf\/iguration space.

\section{Model and landscape}

The system is described by the Hamiltonian 

\setlength{\mathindent}{4cm} 
\begin{equation} \label{ham} H=-\sum_{i<j}J_{ij}S_iS_j \end{equation}

on the simple cubic lattice  with  periodic  boundary conditions. The sum runs over all nearest neighbour pairs of 
Ising spins $S_i$ with values $\pm 1$.
The sample is prepared by randomly assigning exchange couplings $J_{ij}=\pm J$ to the bonds of the lattice. In this paper only one specif\/ic random
arrangement of exchange couplings $\{J_{ij}\}$ for a f\/inite system of the size $N=4\times4\times4$ is used. 
All 1635796 states up to the third excitation were calculated using the branch-and-bound method
of discrete optimisation \cite{hart_d}. The schematic picture of the conf\/iguration space is visualised in Fig. \ref{land}.
It forms a energy landscape consisting of clusters, valleys and barriers. A set of conf\/igurations is called cluster, if a ``chain''
exists connecting them.  The chain is built up by neighbouring  conf\/igurations, where neighbours are states of the same energy, 
which dif\/fer in the orientation of one spin. The landscape is symmetrical due to Eq. (\ref{ham}). Two clusters of dif\/ferent energies are
connected whenever at least one conf\/iguration of the f\/irst cluster dif\/fer from one conf\/iguration of the second cluster by only a one-spin
f\/lip. The two dif\/ferent ground state clusters \#1 and \#2, e.g., consist of 12 and 18 conf\/igurations, respectively. 
Valleys can be assigned to these ground state clusters. A valley puts together all clusters, which only have connections with its ground state cluster.
Dif\/ferent valleys are connected by so-called saddle clusters, which procure the transition over energy barriers.  

\section{Method and dynamics}
\begin{figure}[t]
\centering 
\includegraphics[clip=true,angle=0,scale=0.76]{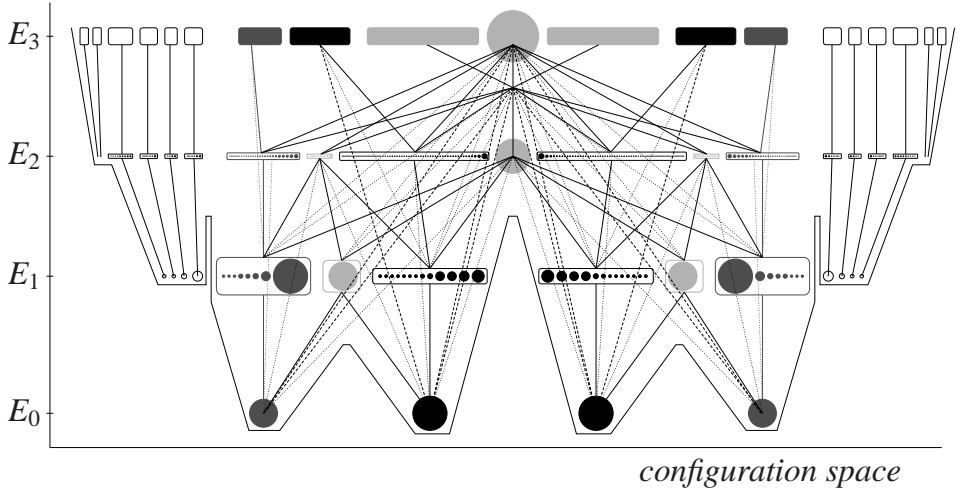}
\caption{\label{land} Schematic picture of the exact low--energy landscape up to the third excitation. Clusters are marked by circles,
 where the size is proportional to the number of configurations in the cluster (note that the scale is dif\/ferent for  dif\/ferent energy levels:
 the largest cluster in the first, second and third excitation contains 819, 82960 and 1503690, respectively).
 The lines denote the one--spin f\/lip connections. All clusters with the same connectivity are pooled by a box.}
\end{figure}

\begin{figure}[b] 
\centering 
\includegraphics[clip=true,angle=0,scale=0.45]{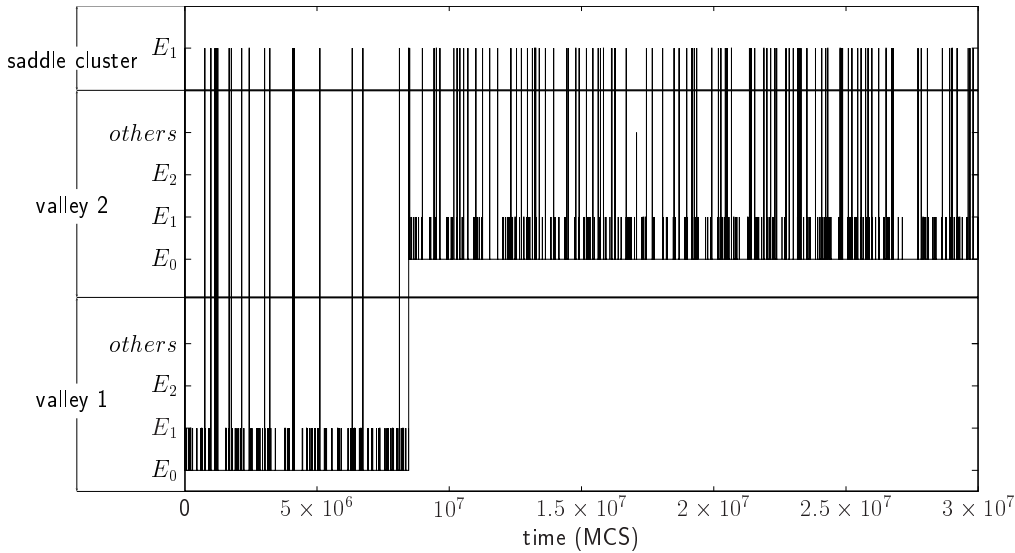}
\caption{\label{run} An individual run through the landscape vs. time   ($\beta=2.5$).}
\end{figure}

\begin{figure}[h]
\centering
\includegraphics[clip=true,angle=0,scale=0.35]{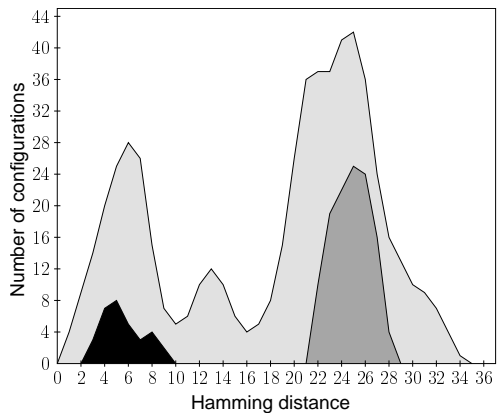}
\caption{\label{prof} The transition prof\/ile of the saddle cluster illustrated by the number of configurations vs. hamming distance from a reference 
state (see text). The shaded area marks all configuration in the saddle cluster.
States having connections with the valley \#1 (dark) and \#2 (middle) are specially emphasised.} 
\end{figure}

\begin{figure}[t]
\centering
\includegraphics[clip=true,angle=-90,scale=0.6]{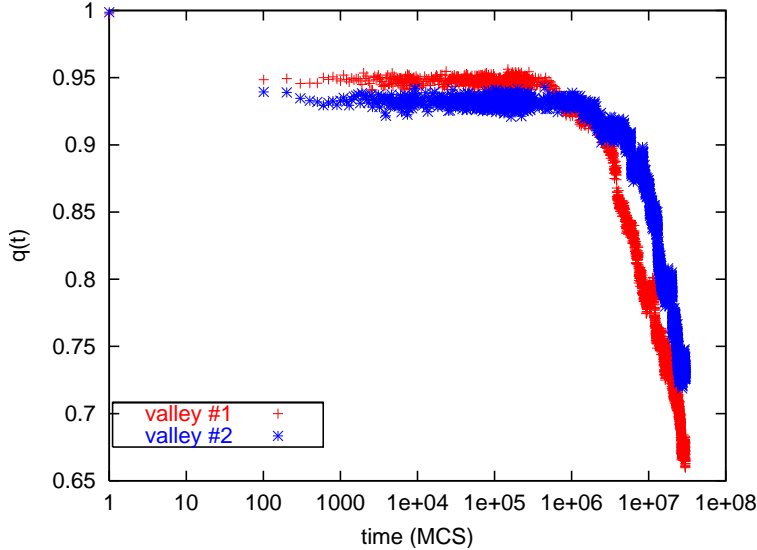} 
\caption{\label{corr} The spin correlation function vs. time. The starting configuration is selected from the set of  ground states of valley
\#1 and \#2 ($\beta=2.5$). }
\end{figure}

The complete knowledge of the low-energy landscape allows us to investigate the inf\/luence of the size and structure of
clusters and valleys and their connectivity on the dynamics.
The time evolution of the system in the conf\/iguration space can be described as the progressive exploration of clusters and valleys.
We use the Monte Carlo Metropolis algorithm  for dif\/ferent $\beta=(k_{B}T)^{-1}$, where $T$ is the temperature of the heat bath  \cite{binder}.
One Monte Carlo step (MCS) is used as time unit. An individual run through the landscape is shown in Fig. \ref{run}.
We  start from  an arbitrary state in the left ground state cluster (\#1) of Fig. \ref{land}. 
First the system  walks in the  valley and sometimes touches  the saddle cluster in the f\/irst excitation. After an escape time $t_{esc}$ 
in the order of $10^7$ MCS the system leaves the f\/irst valley and goes through the saddle cluster to the second one.
This transition is determined by the internal structure of the saddle cluster shown as its transition prof\/ile in Fig. \ref{prof}.
First all pairs of conf\/igurations are checked to f\/ind out the largest hamming distance $h_d$ ($h_d$ of a spin pair is one half of the dif\/ference   
of the sum over all spins).
Then one of these both states is used as reference state and the $h_d$ values of all conf\/igurations of the saddle cluster with respect to 
this reference state are calculated. 
Two sets of states, which have one-spin-f\/lip  connection with the ground states of both af\/f\/iliated valleys, are marked. They denote the input and
the output area for a transition from the f\/irst valley (\#1) to the second one (\#2).  Obviously, a transition as a walking between these sets is slowed 
down due to the smaller numbers of states between.  \\
Quantitatively, the random walk can be described by the spin correlation function
\begin{equation}q(t)=\frac{1}{N}{<S^{G}_{i}(0)S_i(t)>},\end{equation}
where $S^{G}_{i}(0)$ is the $i$-th spin of the  starting conf\/iguration arbitrary chosen from the ground states of valley \#1 (\#2).
The brackets denote the average over 100 runs starting from the same state (Fig. \ref{corr}).

\section{Results and discussion}
The spin correlation function vs. time is characterised by a plateau  with the value  $q_{pl}$ 
followed by a temperature dependent decay. 
To examine the correlation between the structure of the landscape and the dynamics we compare  $q_{pl}$ with
the size of the valley having in mind that the spin correlation within the valley can be calculated 
using the mean hamming distance $\overline h_d$ of all pairs of states by
\begin{equation}\label{coha} q_{pl}^{(ham)}=1-2\overline h_d/N.\end{equation}
We found an agreement between $q_{pl}$ and  $q_{pl}^{(ham)}$ (Table \ref{tabl}), where the average in Eq. (\ref{coha}) is performed
over all states in the ground state cluster. 
\begin{table}[b]
\caption{\label{tabl}{The values of $q_{pl}$ obtained from the simulations (Fig.  \ref{corr}) and calculation (Eq. (\ref{coha}))}}
\begin{center}
\footnotesize
\begin{tabular}{rcccc}\hline
                  & \hspace{0.4cm} &Fig. \ref{corr}    &\hspace{0.5cm} &  Eq. (\ref{coha}) \\ \hline
 $q_{pl}$ (\#1)   &  &$0.947 \pm 0.004$  & &$0.936$  \\ 
 $q_{pl}$ (\#2)   &  &$0.932 \pm 0.004$  & &$0.924$  \\ 
 $\Delta q_{pl}$  &  &$0.015 \pm 0.004$  & & $0.012$ \\ \hline
\end{tabular}
\end{center}
\end{table}
So the plateau ref\/lects the dynamics within the valley. \\
The subsequent decay of $q(t)$ shows the escape from the valley. The escape time $t_{esc}$ depends on the temperature and can be f\/itted by    
$t_{esc} \sim exp(\beta\;\Delta E_{eff})$. We found $\Delta E_{eff}=4.24\pm0.08 \quad (4.46\pm0.09)$ for valley \#1 (\#2), respectively.
Obviously, the ef\/fective energy barrier is larger than the real one, which is $\Delta E= 4$ in our example. Moreover, 
$\Delta E_{eff}$ is larger for the valley \#2 than for \#1. This ref\/lects that the system can leave the saddle cluster easier in direction
to \#2, because there are more exit connections. 

In summary, we have shown that the dynamics of spin glasses is related to the microscopic structure of the energy landscape.
The characteristic shape of the correlation function and the slow dynamics are caused by the restricted connectivity of clusters 
and valleys and their internal prof\/iles. 
Our results are obtained for one particular random set $\{J_{ij}\}$.  
Simulations using dif\/ferent sets conf\/irm that our conclusions are not af\/fected by the choice of $\{J_{ij}\}$.

\begin{ack}
The authors wish to thank A. Heuer for valuable discussions.This work is supported by Graduiertenkolleg 
``Struktur- und Korrelationef\/fekte in Festk\"orpern''. 
\end{ack}






\begin{thebibliography}{7}
 \bibitem{bu_heu} S. B\"uchner, A. Heuer, Phys. Rev. Lett. 84 (2000) 2168.
 \bibitem{Scir} W. Kob, F. Sciortino, P. Tartaglia, Europhys. Lett. 49 (2000) 590.   
 \bibitem{sib_sch} P. Sibani, J.C. Sch\"on, P. Salamon, J.O. Andersson, Europhys. Lett. 22 (1993) 479.
 \bibitem{kl_kobe} T. Klotz, S. Kobe, Acta Phys. Slovaca  44 (1994) 347.
 \bibitem{glt} S.C. Glotzer, N. Jan, P.H. Poole, J. Phys. CM 12 (2000) 6675.
 \bibitem{hart_d} A. Hartwig, F. Daske, S. Kobe, Computer Phys. Commun. 32 (1984) 133.
 \bibitem{binder} K. Binder, D.W. Heermann, Monte Carlo Simulation in Statistical Physics, Springer, Berlin, 1988.   

\end{thebibliography}
\end{document}